\documentclass[twocolumn,secnumarabic,amssymb, amsmath, nofootinbib,tightenlines,
nobibnotes, aps, prl,epsfig]{revtex4}
\usepackage{graphicx}% Include figure files
\usepackage{dcolumn}% Align table columns on decimal point
\usepackage{bm}% bold math
\begin{document}
\preprint{APS/123-QED}
\title{Geometrical scaling behavior of the top structure functions ratio at the LHeC }% Force line breaks with \\

\author{G.R.Boroun}%
 \email{grboroun@gmail.com; boroun@razi.ac.ir }
%\author{B.Rezaei }
%\altaffiliation{brezaei@razi.ac.ir}%Lines break automatically or can be forced with \\
\affiliation{ Physics Department, Razi University, Kermanshah
67149, Iran}% \textbackslash\textbackslash
\date{\today}% It is always \today, today,
\begin{abstract}
%%%%%%%%%%%%%%%%%%%%%%%%%%%%%%%%%%%%%%%%%%%%%%%%%%%%%%%
We consider the ratio of the top structure functions
$R^{t}(\tau_{t})$ in top pair production as a probe of the top
content of the proton at the LHeC project. We study the top
structure functions with the geometrical scaling of gluon
distribution at small $x$ and show that top reduced cross section
exhibits geometrical scaling in a large range of photon
vitualities. This analysis shows that top longitudinal structure
function has sizeable impact on the top reduced cross section at
$Q^{2}{\approx}~ 4m_{t}^{2}$.
\\
%%%%%%%%%%%%%%%%%%%%%%%%%%%%%%%%%%%%%%%%%%%%%%%%%%%%%%%
\end{abstract}
 \pacs{13.60.Hb; 12.38.Bx}%PACS, the Physics and Astronomy
                              %Classification Scheme.
\keywords{Charm Structure Function; Gluon Distribution;
Hard Pomeron; Small-$x$} %Use showkeys class option if keyword
\maketitle
%%%%%%%%%%%%%%%%%%%%%%%%%%%%%%%%%%%%%%%%%%%%%%%%%%%%%%%%%%%%%%%%%
Recently, a method of determination of the top structure function
in the proton from the LHeC project [1-2] has been proposed [3].
On the basis of the method it is known that the dominant source
for the $F_{2}^{t}$ scaling violations is the conversion of gluons
into the $t\overline{t}$ pairs at low-$x$. As the initial scaling
increase as $Q^{2}$ increases from $x_{0}=0.0001-0.1$ according to
$Q^{2}=10-10000~GeV^{2}$ respectively. In this limit, the crucial
point is the observation that the top structure function
parameterization depends directly to the gluon density. The
relevant framework for the dominant of the gluon distributions in
perturbative QCD in this limit is the leading $\log(1/x)$
$({L}L1/x)$ approximation. The basic quantity in this
approximation is the non-integrated gluon distribution
$f(x,k_{T}^{2})$ which is related to the conventional gluon
density $g(x,Q^{2})$, which satisfies DGLAP evolution, as
\begin{equation}
xg(x,Q^{2})=\int^{Q^{2}}_{0}\frac{dk_{T}^{2}}{k_{T}^{2}}f(x,k_{T}^{2}).
\end{equation}
The analytical behavior for $f(x,k_{T}^{2})$ at small $x$ found to
be given by [4], if the running coupling constant effects are
taken into account, as
\begin{eqnarray}
f(x,k_{T}^{2}){\sim}\mathcal{R}(x,k_{T}^{2})x^{-\lambda}
\end{eqnarray}
where $\lambda=4\frac{N_{c}\alpha_{s}}{\pi}\ln(2)$ at LO and at
NLO it has the following form [5]
\begin{equation}
\lambda=4\frac{N_{c}\alpha_{s}}{\pi}\ln(2)[1-c(\frac{1}{2})\frac{N_{c}\alpha_{s}}{4\pi}],
\end{equation}
and
$c(\frac{1}{2})=25.8388+0.1869\frac{n_{f}}{N_{c}}+10.6584\frac{n_{f}}{N_{c}^{3}}$.
The quantity $1+\lambda$ is equal to the intercept of the
so-called BFKL Pomeron. The $K_{T}$-factorization approach relates
strongly to Regge-like behavior of gluon distribution, as we
restrict our investigations to the gluon distribution function at
the following form
\begin{equation}
xg(x,Q^{2}){\sim}x^{-\lambda}.
\end{equation}
Here $\lambda$ is the hard-Pomeron  intercept. The credible
phenomenology intercept of the BFKL equation can be defined by a
kinematic constraint to control of the gluon ladder [4]. The
effect of this constrain on the intercept one finds that
 it reduced the intercept from $\lambda{\sim}0.5$ to
 $\lambda{\sim}0.3$ [6]. Recently the value $0.317$ estimated directly
from the data on the proton unpolrized structure function [7].\\
The latest data [8] for charm and beauty structure functions show
that there are not enough data for the suggestion of the
logarithmic $x$-derivative in the full kinematic range available
as [9]
\begin{eqnarray}
\delta=\frac{{\partial}{\ln}F_{2}^{c,b}}{{\partial}{\ln}\frac{1}{x}}.
\end{eqnarray}
For the charm structure functions, the data points at the values
$12{\leq}Q^{2}{\leq}120~GeV^{2}$ are shown that this derivative is
independent of $x$ for low $x$ values to within the experimental
data and this implies that a power law behavior for charm
structure function as $<\delta>$ is estimated from fits to the H1
data as $<\delta>{\simeq}0.43$. For others $Q^{2}$ values and the
beauty structure functions there are no enough experimental data
that this behavior change in the measured kinematic range. Using
ideas from Regge theory, where gluon distributions have the same
power law behavior (Eq.4) for all H1 experimental data for charm
and beauty structure functions. Our estimations show that
$<\lambda^{c}>{\simeq0.45}\pm^{0.90}_{0.23}$ and
$<\lambda^{b}>{\simeq0.43}\pm^{0.21}_{0.33}$ for charm and beauty
intercepts respectively. These values for $\lambda^{,}$s show that
the hard pomeron behavior [10-13] is dominant. Indeed the hard
pomeron behavior gives a very good description of the data within
the experimental accuracy, not only for the charm structure
function $F_{2}^{c}(x,Q^{2})$, but also for the beauty structure
function $F_{2}^{b}(x,Q^{2})$.\\
In leptoproduction, the primary graph is the Photon-Gluon-Fusion
(PGF) model where the incident virtual photon interacts with a
gluon from the target nucleon for producing $t\overline{t}$ at
leading order (LO) and next-leading-order (NLO) processes at the
LHeC project [1-2] Within the variable-flavor- number scheme
(VFNS). In the LHeC project, we think that the top quark component
$F_{2}^{t}$ of $F_{2}$ is apparently governed almost entirely by
hard-pomeron exchange over a wide range of $x$ and $Q^{2}$. In
LHeC project, for $Q^{2}>2~GeV^{2}$,  the hard pomeron behavior is
driven solely by the gluon field. Therefore, according to
perturbative QCD, the top quark originates from a gluon structure
function that is dominated at small $x$ by hard pomeron exchange.\\
Let us use the gluon distribution to calculate top production in
LO up to NLO pQCD at small $x$ as the top structure functions may
be given by
\begin{eqnarray}
F_{k}^{t}(x,Q^{2},m^{2}_{t})&=&e_{t}^{2}\frac{\alpha_{s}(<\mu^{2}_{t}>)}{2\pi}\int_{1-\frac{1}{a}}^{1-x}dzC_{g,2}^{t}
(1-z,\zeta)\nonumber\\
&& {\times}G(\frac{x}{1-z},<\mu^{2}_{t}>),(k=2~ \& ~L)
\end{eqnarray}
where $C_{g,k}^{t}$ are the coefficients functions and $a = 1+\xi$
where $\xi{\equiv}\frac{m^{2}_{t}}{Q^{2}}$ and $G(=xg)$ is the
gluon momentum distribution. The physical intuition leads us to
take $<\mu^{2}_{t}>=4m_{t}^{2}+Q^{2}/2$ for both, though it must
be recognised that this is a mere guess. The value
$m_{t}=157~GeV$ is fixed for these results.\\
Thus, exploiting the hard pomeron behavior (4) for the gluon
distribution at $x^{-\lambda}{\gg}1$ and using the NLO
approximation for collinear coefficient functions and anomalous
dimensions of Wilson operators. The top structure functions
$F_{k}^{t}$, with respect to the gluon distribution behavior, have
the following forms
\begin{eqnarray}
F_{k}^{t}(x,Q^{2},m^{2}_{t})=e_{t}^{2}\frac{\alpha_{s}(<\mu^{2}>)}{2\pi}\eta_{k}(x,<\mu^{2}>)\nonumber\\
{\times}G(x,<\mu^{2}>),
\end{eqnarray}
where
\begin{eqnarray}
\eta_{k}(x,<\mu^{2}>)=C_{g,k}^{t}(x,\zeta){\otimes}(x)^{\lambda},
\end{eqnarray}
and the symbol $\otimes$ denotes convolution according to the
usual prescription,
$f(x){\otimes}g(x)=\int_{x}^{1}(dy/y)f(y)g(x/y)$. The ratio of the
top structure functions are important for investigation of the
photon-top quark scattering contribution to the  Callan-Gross
ratio at low and moderate $Q^{2}{\simeq}m^{2}_{t}$ as
\begin{eqnarray}
R^{t}(x,Q^{2})=\frac{F_{L}^{t}(x,Q^{2})}{F_{2}^{t}(x,Q^{2})}.
\end{eqnarray}
The solution of Eq.(9) is straightforward and given by
\begin{eqnarray}
R^{t}(x,Q^{2})&=&\frac{e_{t}^{2}\frac{\alpha_{s}(<\mu^{2}>)}{2\pi}\eta_{L}(x,<\mu^{2}>)G(x,<\mu^{2}>)}
{e_{t}^{2}\frac{\alpha_{s}(<\mu^{2}>)}{2\pi}\eta_{2}(x,<\mu^{2}>)G(x,<\mu^{2}>)}\nonumber\\
&&=\frac{\eta_{L}(x,<\mu^{2}>)} {\eta_{2}(x,<\mu^{2}>)}.\nonumber
\end{eqnarray}
In general, we write the quantity $R^{t}$  by the following form
\begin{eqnarray}
R^{t}(x,Q^{2})=\frac{C_{g,L}^{t}(x,\zeta){\otimes}(x)^{\lambda}}{C_{g,2}^{t}(x,\zeta){\otimes}(x)^{\lambda}}.
\end{eqnarray}
In fact, the gluon distribution input cancels in the ratio.
Therefore the reduced cross section for photon- top quark
production [14] is given by
\begin{eqnarray}
\widetilde{\sigma}^{t\overline{t}}(x,Q^{2})=F_{2}^{t}(x,Q^{2})[1-\frac{y^{2}}{1+(1-y)^{2}}\frac{\eta_{L}(x,<\mu^{2}>)}
{\eta_{2}(x,<\mu^{2}>)}],\nonumber\\
\end{eqnarray}
where $y(=\frac{Q^{2}}{sx})$ is the inelasticity variable and $s$
is the square of the center-of- mass energy of the virtual
photon-top quark subprocess $Q^{2}(1-z)/z$.  H1 Collab. [8]
obtained the charm and beauty  structure functions
$F^{c\overline{c}}_{2}$ and $F^{b\overline{b}}_{2}$  from the
measured $c$ and $b$ cross sections after applying small
corrections for the longitudinal structure functions
$F^{c\overline{c}}_{L}$ and $F^{b\overline{b}}_{L}$ at low and
moderate inelasticity. The inelasticity values for $c$ and $b$
production in this experiment were in the region $0.09<y<0.5$. We
expect that inelasticity value for $t$ production to be at high
values of inelasticity. The high $y$ values at the top production
are according to the very low $x$ values, as in this region the
screening (or shadowing) effects are very important. The main
effect of shadowing is the recombining of gluons at higher
densities via the process $gg{\rightarrow}g$, where it causes the
top structure behavior is tamed. The saturation limit for the
gluon distribution is at the order of the hadronic radius $R_{H}$
as $G_{sat}(x,Q^{2}){\sim}R^{2}_{H}Q^{2}/\alpha_{s}(Q^{2})$.
Because the gluons are concentrated around the hot-spots points,
where the radius $R_{hs}$ is smaller than the hadronic radius
$R_{H}$, the linear effects must be modify by the nonlinear terms
as have been formalized by GLRMQ [15].\\
However at low-$y$, where $F_{L}^{t}$ set to zero we have
$\widetilde{\sigma}^{t\overline{t}}(x,Q^{2})({\equiv}\widetilde{\sigma}^{t\overline{t}}_{F_{2}})=F_{2}^{t}(x,Q^{2})$.
But at moderate and high inelasticity, the longitudinal structure
function contributions to the cross section. The fractional
$F_{L}^{t}$ contribution to the top cross section investigate by
\begin{eqnarray}
C_{F_{L}^{t}}{\equiv}\mid
\frac{\widetilde{\sigma}^{t\overline{t}}-\widetilde{\sigma}^{t\overline{t}}_{F_{2}}}{\widetilde{\sigma}^{t\overline{t}}}\mid.
\end{eqnarray}
Indeed, there is a sizeable contribution to the top cross section
at the LHeC project at high $y$ and very low $x$ values. The LHeC
can use a proton beam with energy up to $7~TeV$, and the electron
beam energy is set to $60~GeV$. At fixed $(x, y)$, the gain in
$\sqrt{s}$ will be a factor about 4 as compared to HERA. The
kinematic range of the LHeC for determination of the top structure
function is at low $x$ and at high $Q^{2}$ [16-17]. At small $x$,
the inelasticity is given as $y {\simeq}1- E'_{e}/E_{e}$.
Therefore, we can choose the extremum value for the inelasticity
as if $y{\rightarrow}1$, then $f(y)=y^2/Y_{+}{\rightarrow}1$ where
$Y_{+} =
1+(1-y)^{2}$.\\
Therefore, the $t\overline{t}$-pair production at the LHeC project
in DIS can be happen at small enough $x$ where the geometrical
scaling (SC) has been introduced [18] in this region as the dense
gluon system is fully justified. Thus the saturation scale
$Q_{s}^{2}(x)$, is an intrinsic characteristic this dense gluon
system which tame the rise of the gluon distribution at small $x$.
One thus finds that the saturation scale has the form
$Q_{s}^{2}(x)=Q_{0}^{2}(x/x_{0})^{-\lambda}$ as increases with
decreasing $x$. This type of scaling is also found to be an
intrinsic property of the nonlinear evolution equations. Therefore
the proton cross section is dependence upon the single variable
$\tau=Q^{2}/Q_{s}^{2}(x)$, as
\begin{eqnarray}
\sigma_{\gamma^{*} p}(x,Q^{2})=\sigma_{\gamma^{*} p}(\tau).
\end{eqnarray}
The gluon distribution at the geometric scale is defined by
\begin{eqnarray}
\frac{\alpha_{s}}{2\pi}xg(x,Q^{2}=Q^{2}_{s}(x))=r_{0}x^{-\lambda},
 \end{eqnarray}
with $r_{0}=\frac{3}{8{\pi^{3}}} {\sigma_{0}} {x_{0}^{\lambda}}$.
The two parameters $\sigma_{0}$ and $x_{0}$ determined when
authors in Ref.[19] perform a fit including charm to the total
cross section $\sigma^{\gamma^{*}p}$. Using the leading-twist
relationship between the dipole cross section and the unintegrated
gluon distribution, as the integrated gluon distribution at fixed
coupling is given by [20]
\begin{eqnarray}
G(x,Q^{2})&=&\frac{3\sigma_{0}}{4\pi^{2}\alpha_{s}}(-Q^{2}e^{-Q^{2}(\frac{x}{x_{0}})^{\lambda}}\\\nonumber
&&+(\frac{x_{0}}{x})^{\lambda}(1-e^{-Q^{2}(\frac{x}{x_{0}})^{\lambda}})).
\end{eqnarray}
Therefore we use the same parameters as those were found from a
fit to small $x$ data [19]. But $Q_{0}^{2}$ have to be larger than
$2~GeV^{2}$ and the Bjorken variable $x=x_{B}$ was modified [21]
to be
\begin{eqnarray}
x=x_{B}(1+\frac{4m^{2}_{t}}{Q^{2}}).
\end{eqnarray}
In top production the geometrical scaling violation is expected
due to the large top quark mass, therefore we use the scaling
variable $\tau_{t}$ according to the top quark mass further than
the historically variable [22] as
\begin{eqnarray}
\tau_{t}=(1+\frac{4m_{t}^{2}}{Q^{2}})^{1+\lambda}\frac{Q^{2}}{Q_{0}^{2}}(\frac{x_{B}}{x_{0}})^{\lambda}.
\end{eqnarray}
This new scale is valid in the small $x$ as top pair production is
dominance at this region. Therefore the saturation model leads to
\begin{eqnarray}
F_{k}^{t}(\tau_{t})&=&e_{t}^{2}(\frac{3\sigma_{0}}{8\pi^{3}}(-\mu^{2}e^{-\mu^{2}(\frac{x}{x_{0}})^{\lambda}}\\\nonumber
&&+(\frac{x_{0}}{x})^{\lambda}(1-e^{-\mu^{2}(\frac{x}{x_{0}})^{\lambda}})))\eta_{k}(\tau_{t}).
\end{eqnarray}
Finally the reduced cross section for top pair production in DIS
at the LHeC project is bounded by the geometrical scaling which
assures unitarity of $F_{2}^{t}$ at the limit $y{\rightarrow}1$,
as
\begin{eqnarray}
\widetilde{\sigma}^{t\overline{t}}(\tau_{t}){\rightarrow}F_{2}^{t}(\tau_{t})[1-R^{t}(\tau_{t})].
\end{eqnarray}
In Table I, we find a sizable contribution to the reduced cross
section at high $y$. This overlaps with the high $Q^{2}$ and very
low $x$ region where is outside the kinematic region accessed at
LHeC as $0.000002 < x < 0.8$ and $2 < Q^{2} < 100,000~GeV^{2}$. We
see that the corresponding longitudinal top structure function is
almost zero for $Q^{2}{\leq}1000 GeV^{2}$ at very low $x$ values.
In this case,
$\widetilde{\sigma}^{t\overline{t}}(\tau_{t}){=}F_{2}^{t}(\tau_{t})$.
In Figs.1 and 2, we show the ratio $R^{t}$ in this limit. This
value is non zero for $Q^{2}>1000 GeV^{2} $ and has a maximum
value less than 0.21 practically at $Q^{2}{\simeq}1E6$. Our
results show that the ratio $R^{t}$ is independent of the $x$
values and it has the same behavior for the charm and beauty
production [23-27] in the entire region of $Q^{2}$. We conclude
that the longitudinal top structure function component to the
reduced cross section could be good probe of the top density in
the proton at
$Q^{2}{\simeq}4m_{t}^{2}$.\\
One can also see from Figs.1 and 2 the behavior of the top
structure functions ratio versus the top scaling variable
$\tau_{t}$ for different values of $Q^{2}$. In Fig.3 we show the
top structure functions with $x<1E-3$ for different values of
$Q^{2}$ against the scaling variable $\tau_{t}$. We see that the
results exhibit geometrical scaling over a very board range of
$Q^{2}$ at any $Q^{2}$ scale. We can also clearly see (in Figs.3
and 4) that the behavior of the
$\widetilde{\sigma}^{t\overline{t}}(\tau_{t})$ and
$F_{2}^{t}(\tau_{t})$ on $\tau_{t}$ is  approximately $1/\tau_{t}$
at large $\tau_{t}$. The transition  point is placed at
$\tau{\simeq}0.45$ which has value very less than
$\mu_{t}^{2}=4m_{t}^{2}$ for a top mass $m_{t}=157~GeV$. In this
point the $Q_{s}^{2}$ has value of order $200000~GeV^{2}$, where
in this region $Q^{2}{<<}Q_{s}^{2}$ and the nonlinear effects are
important as the gluon density growth by the rate
$Q_{s}^{2}/\Lambda^{2}$. As plotted in Fig.5, this transition
point will be determined at LHeC project.\\
%%%%%%%%%%%%%%%%%%%%%%%%%%%%%%%%%%%%%%%%%%%%%%%%%%%%%%%%%%%%%%%%%%%%%%%%%
In conclusion, we prediction  the top structure functions at the
LHeC domain with respect to the geometrical scaling. We demonstrated the usefulness the direct extraction
 $F_{2}^{t}$ from the top reduced cross section $\widetilde{{\sigma}}^{t\overline{t}}$ as the top longitudinal structure function has a correlation function
 at $Q^{2}{\geq}4m_{t}^{2}$. Also we show the ratio of the top structure functions as it is independent of $x$ at low $x$ values and it has the same
 behavior as considered for charm and beauty structure function
 ratios. The maximum value estimated for $R^{t}({\tau_{t}})$ is almost
 ${\sim}0.2$ in a wide region of $x$. The most important numerical
 sources of theoretical uncertainty in $t\overline{t}$-pair
 production are the factorization scale dependence and the
 constant parameters in the saturation model. Finally we
 show the geometrical scaling  in the top structure functions from
 the region $x<0.01$ and a transition in the behavior on
 $\tau_{t}$.
 \\

\subsection{Acknowledgment}
 Author is grateful to Prof.B.Kniehl and Prof.N.Armesto for suggestion, Prof. A.Kotikov for reading the manuscript and Prof.N.Ya Ivanov  for reading and useful comments.\\

%%%%%%%%%%%%%%%%%%%%%%%%%%%%%%%%%%%%%%%%%%%%%%%%%%%%%%%%%%%%%%%%%%%%%%%%
\textbf{References}\\
1. P.Newman, Nucl.Phys.Proc.Suppl.{\bf191}, 307(2009);
S.J.BRODSKY, hep-ph/arXiv:1106.5820 (2011); Amanda Cooper-Sarkar,
hep-ph/arXiv:1310.0662 (2013).\\
2. LHeC Study group, CERN-OPEN-2012-015; F. D. Aaron et al. [H1
and ZEUS Collaboration], JHEP {\bf1001},
109(2010)[hep-ex/arXiv:0911.0884
]; LHeC Study group, LHeC-Note-2012-005 GEN.\\
3. G.R.Boroun, [hep-ph/arXiv:1411.6492],  Phys.Lett.B{\bf 741}, 197(2015).\\
4.J.Kwiecinski, J.Phys.G{\bf 19}, 1443(1993); J.Kwiecinski, A.D.Martin, R.Roberts and W.J.Stirling, Phys.Rev.D{\bf 42}, 3645(1990).\\
5. V.S.Fadin and L.N.Lipatov, Phys. Lett. B{\bf 429}, 127(1998).\\
6.A.M.Cooper-Sarkar and R.C.E.Devenish, Acta.Phys.Polon.B{\bf 34}, 2911(2003).\\
7. A.A.Godizov, Nucl.Phys.A{\bf 927}, 36(2014).\\
8. F.D. Aaron et al. [H1 Collaboration],
Eur.Phys.J.C\textbf{65},89(2010).\\
9. A.Kotikov,  hep-ph/arXiv:1212.3733 (2012).\\
10. A. Donnachie and  P. V. Landshoff, Phys. Lett. B{\bf 296},
257(1992); A. Donnachie and P. V. Landshoff, Phys. Lett. B{\bf 437}, 408(1998 ).\\
11. A. Donnachie and P. V. Landshoff, Phys. Lett. B{\bf 550}, 160(2002 ); Phys. Lett. B{\bf 533}, 277(2002); Phys. Lett. B{\bf 595}, 393(2004).\\
12. P.V.Landshoff, arXiv:hep-ph/0203084(2002).\\
13. J.R.Cudell and G.Soyez, Phys. Lett. B{\bf 516}, 77(2001);
J.R.Cudell, A. Donnachie and P. V. Landshoff, Phys. Lett. B{\bf
448},
281(1999).\\
14. A.Y.Illarionov, B.A.Kniehl and A.V.Kotikov, Phys.Lett.B {\bf 663}, 66(2008); A. Y. Illarionov and A. V. Kotikov, Phys.Atom.Nucl. {\bf75}, 1234(2012).\\
15. A.H.Mueller and J.Qiu, Nucl.Phys.B\textbf{268}, 427(1986);
L.V.Gribov, E.M.Levin and M.G.Ryskin, Phys.Rep.\textbf{100},
 1(1983).\\
16. Workshop on the LHeC, 20-21 January (2014),
Chavannes-de-Bogis, Switzerland, http://cern.ch/lhec.\\
17. J.L.Abelleira Fernandez, et.al., [LHeC group],
arXiv:1206.2913v1 [physics.acc-ph] 13 Jun (2012).\\
18. A.M.Stasto, K.J.Golec-Biernat, J.Kwiecinski,
Phys.Rev.Lett.{\bf86}, 596(2001).\\
19. K. Golec-Biernat and M.Wusthoff, Phys.Rev.D\textbf{59},
014017(1998).\\
20.R.S.Thorne, Phys.Rev.D{\bf71},054024(2005).\\
21. G.Beuf, C.Royon and D.Salek, arXiv:hep-ph/0810.5082(2008).\\
22. T.Stebel, arXiv:hep-ph/1305.2583(2013).\\
23. N.N.Nikolaev and V.R.Zoller, Phys.Atom.Nucl\textbf{73},
672(2010); Phys.Lett.B \textbf{509}, 283(2001);  N.N.Nikolaev,
J.Speth and V.R.Zoller, Phys.Lett.B\textbf{473}, 157(2000);
R.Fiore,
N.N.Nikolaev and V.R.Zoller, JETP Lett\textbf{90}, 319(2009).\\
24.A.V.Kotikov, A.V.Lipatov, G.Parente and N.P.Zotov, Eur.\
Phys.\ J.\  C {\bf 26}, 51 (2002).\\
25. N.Ya.Ivanov, and B.A.Kniehl, Eur.Phys.J.C\textbf{59}, 647(2009);  N.Ya.Ivanov, Nucl.Phys.B\textbf{814}, 142(2009); Eur.Phys.J.C{\bf59}, 647(2009)\\
26. I.P.Ivanov and N.Nikolaev,Phys.Rev.D{\bf65},054004(2002).\\
27. N.N.Nikolaev and V.R.Zoller, Phys.Lett. B\textbf{509},
283(2001); Phys.Atom.Nucl.\textbf{73}, 672(2010); V.R.Zoller,
Phys.Lett. B\textbf{509},
69(2001).\\

%%%%%%%%%%%%%%%%%%%%%%%%%%%%%%%%%%%%%%%%%%%%%%%%%%%%%%%%%%%%%%%%%%%%%%%%%%%%%%%%
%%%%%%%%%%%%%%%%%%%%%%%%%%%%%%%%%%%%%%%%%%%%%%%%%%%%%%%%%%
\begin{table}
\centering \caption{ The fractional $F_{L}^{t}$ contribution to
the top reduced cross section in bins for top production at the
LHeC. }\label{table:table1}
\begin{minipage}{\linewidth}
\renewcommand{\thefootnote}{\thempfootnote}
\centering
\begin{tabular}{|l|c|} \hline\noalign{\smallskip} $Q^{2}(GeV^{2})$ & $ C_{F_{L}^{t}}$  \\
\hline\noalign{\smallskip}
10 & 0.27E-4 \\
100 & 0.27E-3 \\
1000 & 0.27E-2 \\
10000 & 0.025 \\
100000 & 0.150 \\
1000000 & 0.253 \\
10000000 & 0.200 \\
100000000 & 0.146 \\
1000000000 & 0.113 \\
10000000000 & 0.092 \\
\hline\noalign{\smallskip}
\end{tabular}
\end{minipage}
\end{table}

%%%%%%%%%%%%%%%%%
\begin{figure}
\includegraphics[width=0.5\textwidth]{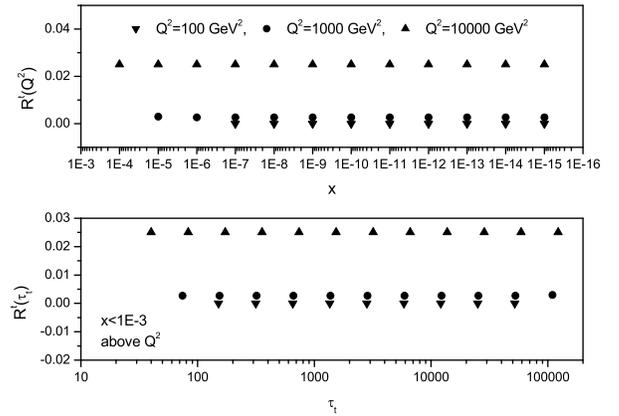}
\caption{ The ratio $R^{t}$ as a function of $x$ and $\tau_{t}$
for different values of $Q^{2}$. }\label{Fig1}
\end{figure}
\begin{figure}
\includegraphics[width=0.5\textwidth]{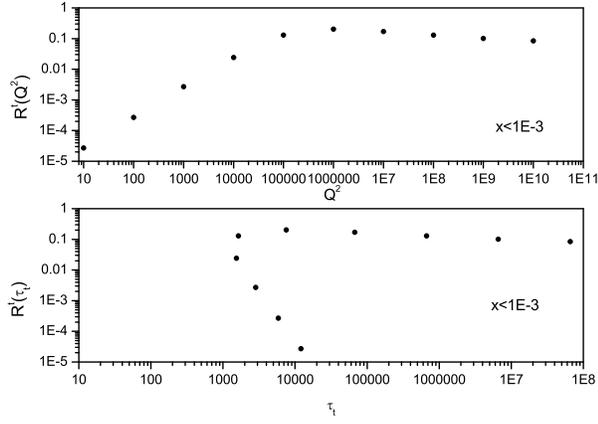}
\caption{ The ratio $R^{t}$ as a function of $Q^{2}$ and
$\tau_{t}$ with $<\mu^{2}>=4m_{t}^{2}+{Q^{2}}/{2}$ . }\label{Fig1}
\end{figure}
\begin{figure}
\includegraphics[width=0.5\textwidth]{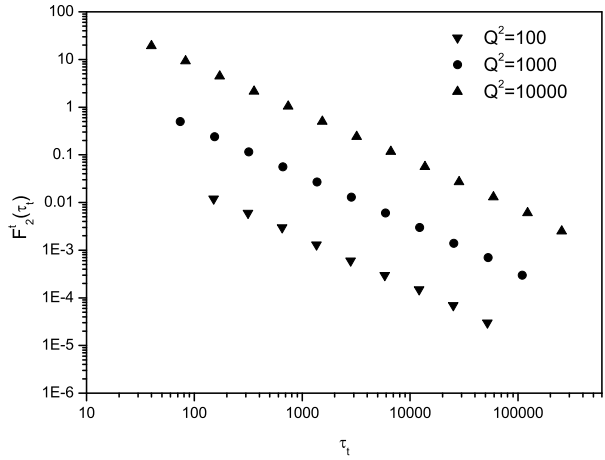}
\caption{ The $F_{2}^{t}$ structure function for
$Q^{2}<4m^{2}_{t}$. }\label{Fig1}
\end{figure}
\begin{figure}
\includegraphics[width=0.5\textwidth]{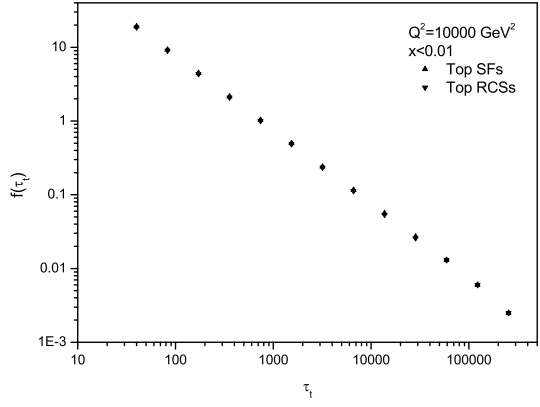}
\caption{ $\widetilde{{\sigma}}^{t\overline{t}}$ (Top RCSs) and
$F_{2}^{t}$ (Top SFs) plotted versus the top scaling variable
$\tau_{t}$. }\label{Fig1}
\end{figure}
\begin{figure}
\includegraphics[width=0.5\textwidth]{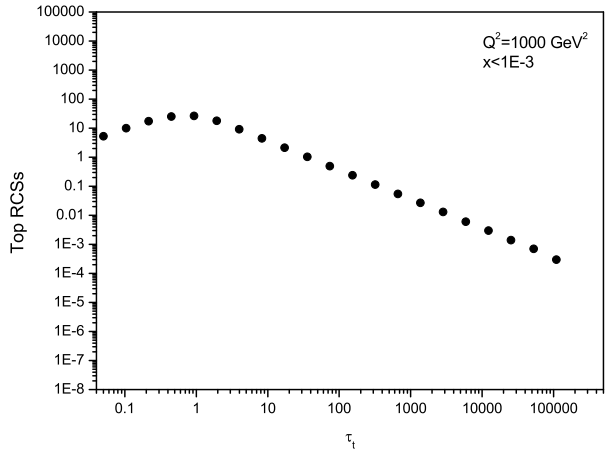}
\caption{ The top reduced cross section
$\widetilde{{\sigma}}^{t\overline{t}}$ from the region $x<0.001$
plotted versus the top scaling variable $\tau_{t}$. }\label{Fig1}
\end{figure}
%%%%%%%%%%%%%%%%%%%%%%%%%%%%%%%%%%%%%%%%%%%%%%%%%%%%%%%%
\end{document}